\documentstyle[prl,aps,graphicx,multicol]{revtex}
  \renewcommand{\narrowtext}{\begin{multicols}{2} \global\columnwidth20.5pc}
  \renewcommand{\widetext}{\end{multicols} \global\columnwidth42.5pc}

\begin{document}
%\twocolumn[\hsize\textwidth\columnwidth\hsize\csname@twocolumnfalse\endcsname

\title{Mechanical Cooper pair transportation as a source of long distance superconducting phase coherence.}
\author{A. Isacsson$^1$, L. Y. Gorelik$^1$, R. I. Shekhter$^1$, Y. M. Galperin$^{2,3}$ and M. Jonson$^1$}
\address{$^1$Department of Applied Physics, Chalmers University of
Technology, SE-412 96 G\"oteborg, Sweden.\\
$^2$Department of Physics, University of Oslo, P. O. Box 1048, N-0316 Oslo, Norway.\\
$^3$Division of Solid State Physics, Ioffe Institute of the Russian Academy of Sciences, St. Petersburg 194021, Russia.}
\date{\today}
\maketitle

\begin{abstract}
Transportation of Cooper-pairs by a movable single Cooper-pair-box placed 
between two remote superconductors 
is shown to establish coherent coupling between them. 
This coupling is due to entanglement of the movable box with the leads and is
manifested in the supression of quantum fluctuations 
of the relative phase of the order parameters of the leads.
It can be probed by attaching a high resistance Josephson junction between the leads and measuring the
current through this junction.
The current is suppressed with increasing temperature.

\end{abstract}
\vspace*{0.1in}
\pacs{PACS number: ????}
\narrowtext
Delocalization of electrons due to tunneling between two bodies
constituting a system will, according to the Heisenberg uncertainty principle, lower
the total energy of this system and, hence, lead to 
interbody coupling. If the bodies are superconductors, charge exchange
due to Cooper pair tunneling (Josephson tunneling) causes such coupling.  
Since the Cooper-pair exchange is sensitive to the relative phases of the 
superconducting order parameters of the bodies this coupling leads to the establishment
of phase coherence i.e. a {\it phase ordering} of the relative phases occurs.
The characteristic energy responsible for this ordering is the
Josephson coupling energy $E_J$.

If charge exchange is accompanied by charge accumulation Coulomb energy 
cannot be neglected. In nanostructures, where the excess charge can be 
strongly confined, this may result in forces strong enough to 
cause mechanical deformations of the systems. Coupling
between mechanical degrees of freedom and electronic degrees of freedom
can then arise in systems having a rigidity comparable to these Coulomb forces. 

That such electromechanical coupling can lead to 
mechanical transportation of charge in a movable nanocluster was suggested
in \cite{chalmers_shuttle} as an unusual mechanism for shuttling of 
electric charge between two bodies (see also \cite{physicaB,cantilever_shuttle}). 
This triggered a number of both 
experimental \cite{Blick,Touminen,Park} and theoretical
\cite{Weiss,Schoeller} activities including 
\cite{nature_paper} where the possibility of coherent transfer of Cooper pairs by
a movable superconducting island was proposed. 

There, a non stationary quantum mechanical process involving
a movable qubit, a coherent superposition of differently charged 
states on a movable superconducting grain, was shown to be responsible for a 
finite supercurrent between two remote phase coherent superconductors.
However, when confronted with this result one is faced by a most fundamental question:
{\it Could such a non-equilibrium quantum mechanical process serve as a source
for the creation of phase coherence (phase ordering) if the two 
superconductors were initially in states with definite number of particles?}  
This paper gives a positive answer to the above question.
\begin{figure}
\centerline{
\includegraphics[width=5cm]{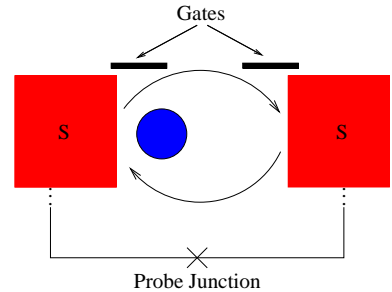}
}
\vspace*{0.5cm}
\caption{Schematic diagram of the system discussed in the text. A movable nanoscale superconducting grain 
which may accomodate excess charge carried by either 0 or 1 extra Cooper pairs
is placed between two finite isolated superconducting leads. Close to the leads,
electrostatic gates ensure that the Coulomb blockade of tunneling is lifted.
After a sequence of grain rotations, the grain moving periodically contacting
the two leads sequentially, a probe (Josephson) junction is connected through
which a current $I$ is measured.}
\label{f_fig1}
\end{figure}
To demonstrate this, we consider the same system as 
in \cite{nature_paper} but with the low impedance 
link connecting the two leads removed. This is shown in Fig.~\ref{f_fig1}. 
Initially the two superconducting leads, labelled left (L) and right (R), are in 
projected BCS-states $\left|N_L\right>$ and $\left|N_R\right>$ with $N_L$ and $N_R$ 
extra Cooper-pairs respectively. In terms of BCS-states $\left|\phi\right>$ with 
definite phase $\phi$ we have
$$\left|N_{L,R}\right>\equiv\frac{1}{\sqrt{2\pi}}\int\limits_{0}^{2\pi}d\phi 
e^{-iN_{L,R}\phi}\left|\phi\right>.$$ 
Transportation of Cooper-pairs between the leads is then provided by a
movable nanometer-sized grain~\cite{crit} placed between the two 
electrodes. When the grain is close to the superconductors electrostatic gates 
ensure that there exists two nearby charge states, differing by one Cooper pair on the grain, 
for which the difference in charging energy $\Delta E_C$ is much smaller than the Josephson energy $E_J$.
Due to the smallness of the grain, Coulomb blockade of tunneling will
restrict the available states of the grain to only these two states 
which we label $\left|n=0\right>$ and $\left|n=1\right>$, 
i.e., a single-Cooper-pair box situation is realized close to each lead.
In this situation, when $E_J\gg \Delta E_C$, a 
coherent superposition, $\alpha\left|0\right>+\beta\left|1\right>$, of differently 
charged states can be created on the grain. This was recently demonstrated 
experimentally\cite{nakamura}.

Consider now the situation when the system is initially in a pure state with 
charge neutral components 
$\left|\Psi\right>=\left|n=0\right>\left|N_L=0\right>\left|N_R=0\right>$. 
If the grain executes periodic motion, performing repeated alternating contacts with the 
two leads, Cooper-pair exchange between the leads will result. Thus the initial state 
will evolve into a superposition of states with different numbers of particles
on the leads,
\begin{eqnarray}
& &\left|n=0\right>\left|N_L=0\right>\left|N_R=0\right>\rightarrow\nonumber \\
& &\sum\limits_{n=0,1}
\sum\limits_{N_L}\sum\limits_{N_R}c_{{N_L},n}\delta_{n+N_L+n_R,0}\left|n\right>
\left|N_L\right>\left|N_R\right>. \nonumber
\end{eqnarray}
The Kroenecker delta makes sure that the total number of particles is conserved. 
To investigate whether this corresponds to a {\it phase ordered state} with 
a well defined phase difference, $\Delta\Phi\equiv\Phi_R-\Phi_L$, 
between the leads a small probe junction,
a Josephson weak link, is connected as shown in Fig.~1.
This provides the means for measuring such order 
since the average current $I$ through this link is related to $\Delta\Phi$,
$$I={\rm Tr}\rho\hat{I},\mbox{ } \hat{I}\equiv I_c\sin(\hat{\Phi}_R-\hat{\Phi}_L).$$
Here $\rho$ is the system density matrix, $I_c$ is the critical current
of the probe junction and $\hat\Phi_{L,R}$ are the phase operators of the respective 
lead. 

To be more concrete, consider the following Hamiltonian describing the system dynamics  
$$H_0(x(t))=-\sum\limits_{s=L,R}E_J^s(x(t))\cos(\hat\Phi_s-\hat\phi)+\Delta E_C(x(t))\hat{n}.$$
$H_0$ depends on time through the grain position $x(t)$.
The first term represents the coupling between the grain and the leads in terms of the 
phase operators of the leads, $\hat\Phi_{L,R}$, and the grain, $\hat\phi$, and time 
varying Josephson coupling energies $E_J^{L,R}(x(t))$. The second term, containing the 
operator $\hat{n}$ for the grain, accounts for the difference in charging 
energy, $\Delta E_C$,
associated with different charge states when the grain is moving between the leads.
In order for a phase ordering to occur two additional ingredients are required:
First, in its present form $[H_0,\hat{I}]=0$ which means that no sign of any ordering 
can be detected. Second, a term describing dissipation/decoherence is needed in order 
for the system to reach a well defined final state irrespective of initial conditions.
If one takes into account that any real isolated superconductor has a finite capacitance
the first point no longer holds. To mimic the behavior when the leads have finite 
capacitance we restrict the Hilbert space of the leads to states $\left|N_{L,R}=m\right>$ 
with $|m|\leq N$.     
To include the effect of the environment we consider the effect of the coupling between the
charge on the grain and the fluctuations of the gate voltage through the term
$$V=\epsilon(2e\hat{n})\sum\limits_{\alpha=1}^{M}C_\alpha\hat{x}_\alpha,\mbox{ } 0\leq \epsilon \leq 1,$$
where $\epsilon$ is a dimensionless coupling constant\cite{Ingold}.
The $\hat{x}_\alpha$ are the coordinates of a harmonic oscillator bath  
$$H_{\cal E}=\sum\limits_{\alpha}\left(\frac{\hat{p}_\alpha^2}{2m_\alpha}
+\frac{1}{2}m_\alpha\omega_\alpha^2\hat{x}_\alpha^2\right)$$
characterized by a spectral density 
$$J(\omega)\equiv\frac{\pi}{2}\sum\limits_{\alpha}
\frac{C_\alpha^2}{m_\alpha\omega_\alpha}\delta(\omega-\omega_\alpha)=\frac{\hbar\omega R}{1+\omega^2 R^2C^2}$$
where $R$ is the resistance and $C$ the capacitance as seen from the gate. 
Hence, the Hamiltonian describing system, environment and interaction between the two is
$$H(t)=H_0(t)+H_{\cal E}+V.$$

Writing the Liouville-von-Neumann equation
in the interaction representation, 
$$\frac{d\hat{\rho}(t)}{dt}=-\frac{i}{\hbar}[\hat{V}(t),\hat{\rho}(t)],$$
$V$ can be treated perturbatively to second order. Tracing out 
the environmental states yields for the reduced density matrix of the system, 
$\hat{\rho}_S(t)\equiv{\rm Tr}_{\cal E}\hat\rho(t)$, 
\begin{eqnarray}
\hat{\rho}_S(t)&=&\hat{\rho}_S(t_0)\nonumber \\
&-&\int\limits_{t_0}^tdt_1 [\hat{n}(t_1),
\int\limits_{t_0}^{t_1}dt_2 K_1(t_1-t_2)[\hat{n}(t_2),\hat{\rho}_S(t_0)]]\nonumber\\
&+&i\int\limits_{t_0}^tdt_1 [\hat{n}(t_1),
\int\limits_{t_0}^{t_1}dt_2 K_2(t_1-t_2)\{\hat{n}(t_2),\hat{\rho}_S(t_0)\}]\nonumber \\
&=&\hat{\rho}_S(t_0)+\hat{\cal L}[\hat{\rho}(t_0)](t).
\label{evolution}
\end{eqnarray}
The kernels $K_{1,2}$ appearing here are related to the spectral density and the 
inverse temperature $\beta=(k_BT)^{-1}$ through
$$K_1(t)=\frac{4e^2\epsilon^2}{\pi\hbar^2}\int\limits_0^{+\infty}d\omega J(\omega)\cos\omega t 
\coth\frac{\beta\hbar\omega}{2}$$
and 
$$K_2(t)=\frac{4e^2\epsilon^2}{\pi\hbar^2}\int\limits_0^{+\infty}d\omega J(\omega)\sin\omega t.$$
To find the (quasi) stationary state of the system eq.(\ref{evolution}) is iterated. To be more precise,
suppose the grain performs rotations with a period $4T_0$. Introduce the following time labels
\begin{itemize}
\item $t_0$: The grain leaves the left electrode.
\item $t_1$: The grain arrives at the right electrode.
\item $t_2$: The grain leaves the right electrode.
\item $t_3$: The grain arrives at the left electrode.
\item $t_4=t_0+4T_0$ the grain leaves the left electrode.
\end{itemize}
The reduced density matrix after a complete rotation is then found by (numerical) iteration, i.e. 
$$\hat{\rho}_S(t_{i+1})=\hat{\rho}_S(t_{i})+\hat{\cal L}[\hat{\rho}(t_{i})](t_{i+1}).$$
For the algorithm to converge the coupling to the bath must be weak enough. The relevant strength
of the perturbation is characterized by the factor $\epsilon^2\eta$ where 
$\eta\equiv(4e^2T_0)/(\pi\hbar C)=4T_0/R_QC$. 
%The steady state probe junction current obtained after iteration
%is shown as a function of the perturbation strength $\epsilon^2\eta$ in Fig.~\ref{pert_strength} over a region where the algorithm is convergent.
%Two tinghs are to be noted. First, in the limit $\epsilon^2\eta\rightarrow 0^+$, 
%when lowest order perturbation theory becomes exact, the current tends to a finite nonzero value.
%Second, the current increases with increasing coupling strength instead of decreasing. The latter a result
%of using finite order perturbation theory. Rather we expect the exact solution to behave as (the guessed)
%dashed line.  
%\begin{figure}
%\centerline{
%\includegraphics[width=7cm]{convalido.eps}
%}\vspace*{0.5cm}
%\caption{ Probe junction current versus perturbation strength. The solid line represents
%the result of the perturbative method used. The dashed line shows the expected behavior of the exact result.}\label{pert_strength}
%\end{figure}

Inspired by \cite{Blick} we consider the grain oscillating in such a way that
it contacts each of the leads for $T_0=1$ ns and that it takes the same time
for it cross the gap between the leads. 
Further, a reasonable value of the Josephson energy during the contacts is\cite{nakamura}
$E_J=7\mu eV$ and we will assume that $\Delta E_C$ is of the same order. For simplicity we
take the time dependence of $E_J$ and $\Delta E_C$ to be step like. Thus
close to the leads $E_J$ takes a constant value while $\Delta E_C$ is zero and vice versa.
More relevant than the exact value of $\Delta E_C$ are the {\it dynamical phase differences} 
$\chi_\pm$ picked up
by the grain during the motion between the leads. More precisely we define
$$\chi_+=\hbar^{-1}\int\limits_{t_0}^{t_1}dt \Delta E_C(t), 
\mbox{ }\chi_-=\hbar^{-1}\int\limits_{t_2}^{t_3}dt \Delta E_C(t).$$

\begin{figure}
\centerline{
\includegraphics[width=7cm]{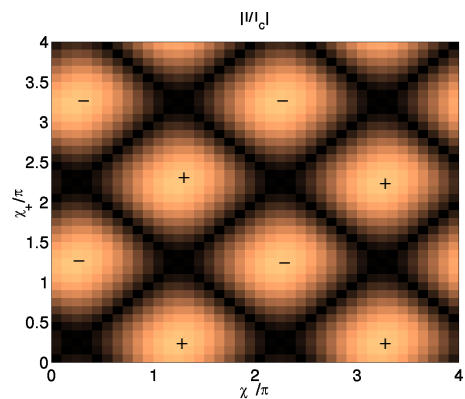}
}\vspace*{0.5cm}
\caption{Average current $I$ through a probe juction attached as in Fig.~\ref{f_fig1} after many grain rotations. 
The magnitude of the current is shown as a function of the phases 
$\chi_\pm$ (in units of $\pi$). Bright areas correspond to large current while $+$ and $-$ indicate its direction.}\label{f_fig2}
\end{figure}

\begin{figure}
\centerline{
\includegraphics[width=7cm]{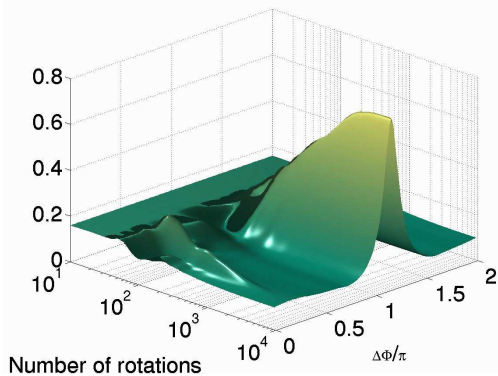}
}\vspace*{0.5cm}
\caption{ Phase difference probability density $f(\Delta\Phi)$ as a 
function of the number of grain rotations.
The graph shows how the system evolves from a state with maximum to a state with minimum 
quantum fluctuations as the number of rotations increases.}\label{f_fig3}
\end{figure}

The dependence of the probe junction current on the dynamical phase difference is shown in
figure~\ref{f_fig2}. Here the maximum capacity of the leads is $N=8$ and the temperature 
set to 1 mK. Starting from an initially pure state the current stabilizes to
a fixed value after a large number of rotations. This current shows a distinct oscillatory 
dependence on the dynamical phases $\chi_\pm$. The number of rotations required before the system
stabilizes depends on the strength of the coupling to the bath. Since our approach is perturbative
a very weak coupling has been used in order to obtain the numerical results. This implies in our case, as is evident 
from Fig.~\ref{f_fig3}, that approximately $10^4$ rotations are neccessary for the value 
of the probe junction current to stabilize. In a real situation this time may be considerably shorter.  

Although a finite current will flow through the weak link in the probe junction it 
is not immediately clear
that the state built up by the rotations is a BCS-type state. 
To see that this is indeed the case consider the
BCS-type phase-states 
$$\left|\phi\right>=\frac{1}{\sqrt{2\pi}}\sum\limits_n e^{in\phi}\left|n\right>.$$
Any state of the system can then be represented in the basis of the phases of the leads 
$\left|\Phi_{L,R}\right>$ and the grain $\left|\phi\right>$ but also in terms of the 
relative phases $\Delta\Phi=\Phi_R-\Phi_L$ and $\Delta\phi=\Phi_R-\phi$
\begin{eqnarray}
&&\left|\Delta\Phi,\Delta\phi\right>=\nonumber\\
&\mbox{ }&\frac{1}{2\pi}\sum\limits_{n=0,1}\sum\limits_{N_L=-N}^{N-n}
e^{-iN_L\Delta\Phi}e^{-in\Delta\phi}\left|n\right>\left|N_L\right>
\left|-N_L-n\right>.\nonumber
\end{eqnarray}
Since the relative phase $\Delta\phi$ between the grain and the right lead necessarily 
suffers from large quantum fluctuations we define
the {\it 'phase difference probability density'} $f(\Delta\Phi)$  as the average 
$$f(\Delta\Phi)\equiv\int\limits_0^{2\pi} d(\Delta\phi)
\left<\Delta\Phi,\Delta\phi\right|\rho\left|\Delta\Phi,\Delta\phi\right>\nonumber $$
In Fig.~\ref{f_fig3} $f(\Delta\Phi)$ has been calculated as a function of the number 
of rotations performed for a system with $N=20$ instead of $N=8$. In this simulation 
the fact that a definite phase relation is built up is numerically verified.
In the beginning of the simulation when the system is in a pure state with 
definite number of particles on each of the
individual components the phase uncertainty is maximal, i.e. 
$f(\Delta\Phi)=(2\pi)^{-1}$. As the
system approaches a stable regime a sharp peak appears around a 
well defined value of $\Delta\Phi$. The location of the peak then defines 
the phase difference between the two leads. 
%Finally we demonstrate that an effective coupling energy, $E_{\rm eff.}$ can be defined.
The exact value of the built up phase difference depends on the parameters of the system. Especially
it depends on the phases $\chi_\pm$ but also on the time spent by the grain in contact with the leads.
In general, the final distribution $f(\Delta\Phi)$ has two peaks separated by a distance $\pi$. The 
relative magnitude of the peaks vary.  For instance, the black lines, representing zero current, 
in Fig.~\ref{f_fig2} corresponding to the lines $\chi_++\chi_-=constant$ are due to a superposition of 
two equally strong built up phases differing by $\pi$ and hence carrying an equal amount of current in opposite directions.
A remnant of such a superposition can bee seen in the distribution in Fig.~\ref{f_fig3}. Here, however, 
only one of the two states is stable while the other one rapidly decays. The largest magnitude 
of the current is obtained when one of the two peaks is completely suppressed. We note further, 
that it is not possible to obtain the value of the built up phase difference by equating the 
final expression for the current in~\cite{nature_paper} to zero since that
expression was derived under the explicit assumption that the phases were fixed.

\begin{figure}
\centerline{
\includegraphics[width=5.7cm]{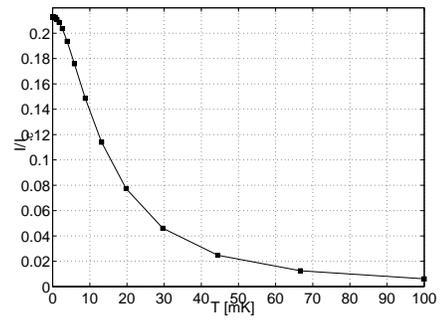}
}\vspace*{0.5cm}
\caption{ Probe junction current versus temperature. The probe 
junction current is here shown for fixed dynamical phases for a series 
of different temperatures. }\label{f_fig5}
\end{figure}

By varying the temperature, $T$, in the simulations the
current $I(T)$ through the probe junction was calculated. In Fig.~\ref{f_fig5} 
$I(T)/I_c$ is shown as a function of temperature. The decrease in current with 
increasing temperature is due to an increased width of the phase distribution. 
At very low temperatures the current saturates due to the fact that quantum fluctuations 
dominate over the thermal fluctuations. Then, with increasing temperature,
 the current decays exponentially at first but with a crossover to algebraic decay for high temperatures.  
%It is clear that there exist a range where $I(T)\sim \exp(-k_BT/E)$
%where $E\approx 30$ mK. It is still an open question whether this energy provides a measure of 
%the strength of the induced coupling.

In conclusion we have shown that mechanically assisted transportation of Cooper
pairs between initially uncoupled, remote superconductors leads to a supression
of quantum fluctuations of the relative phase of their order parameters. 
The movable single Cooper-pair-box, (which may be in a superposition
of two charge states (qubit)) executing sequential
tunneling contacts with each of the superconductors, is responsible for creating the strong entaglement
causing coherent coupling. This coupling is supressed 
with increasing temperatures.

We acknowledge financial support from the Swedish SSF through the programmes 
QDNS (A. I.) and ATOMICS (L. Y. G.) and from the Swedish VR (R. I. S.).
 
\widetext
\end{document}